\documentclass[a4paper,12pt]{article}

\usepackage{lmodern}
\usepackage[T1]{fontenc}
\usepackage[utf8]{inputenc}
\usepackage{textcomp}
\usepackage[french,english]{babel}
\usepackage{amssymb,amsmath,latexsym}
\usepackage{natbib}
\usepackage{graphicx}
\usepackage{booktabs}
\usepackage{multirow}
\usepackage{colortbl}
\usepackage{rotating}
\usepackage[font=it]{caption}
\usepackage{hyperref}
\usepackage{verbatim}

\newtheorem{theorem}{Theorem}

\begin{document}

\newpage
\setcounter{page}{1}

\title{Checking account activity and credit default risk of enterprises: An application of statistical learning methods}
\author{Jinglun YAO\footnote{Student at Ecole Polytechnique}, Maxime LEVY-CHAPIRA\footnote{Quantitative Risk Project Manager at Société Générale}, Mamikon MARGARYAN\footnote{Head of Credit Risk Modeling at Société Générale}}
\date{\today}
\maketitle
\selectlanguage{english}
\begin{abstract}
  The existence of asymmetric information has always been a major concern for financial institutions. Financial intermediaries such as commercial banks need to study the quality of potential borrowers in order to make their decision on corporate loans. Classical methods model the default probability by financial ratios using the logistic regression. As one of the major commercial banks in France, we have access to the the account activities of corporate clients. We show that this transactional data outperforms classical financial ratios in predicting the default event. As the new data reflects the real time status of cash flow, this result confirms our intuition that liquidity plays an important role in the phenomenon of default. Moreover, the two data sets are supplementary to each other to a certain extent: the merged data has a better prediction power than each individual data. We have adopted some advanced machine learning methods and analyzed their characteristics. The correct use of these methods helps us to acquire a deeper understanding of the role of central factors in the phenomenon of default, such as credit line violations and cash inflows.
\end{abstract}
\selectlanguage{french}
\begin{abstract}
  L'existence de l'asymétrie de l'information est une problématique majeure pour les institutions financières. Les intermédiaires financiers, telles que banques commerciales, doivent étudier la qualité des emprunteurs potentiels afin de prendre leurs décisions sur les prêts commerciaux. Les méthodes classiques modélisent la probabilité de défaut par les ratios financiers en utilisant la régression logistique. Au sein d'une principale banque commerciale en France, nous avons accès aux informations sur les activités du compte des clients commerciaux. Nous montrons que les données transactionnelles surperforment les ratios financiers sur la prédiction du défaut. Comme ces nouvelles données reflètent le flux de trésorerie en temps réel, ce résultat confirme notre intuition que la liquidité joue un rôle essentiel dans les phénomènes de défault. En outre, les deux bases de données sont complémentaires l'une à l'autre d'une certaine mesure: la base fusionnée a une meilleure performance de prédiction que chaque base individuelle. Nous avons adopté plusieurs méthodes avancées de l'apprentissage statistique et analysé leurs caractéristiques. L'utilisation appropriée de ces méthodes nous aide à acquérir une compréhension profonde du rôle des facteurs centraux dans la prédiction du défaut, tel que la violation de l'autorisation du découvert et les flux de trésorerie.
\end{abstract}

\selectlanguage{english}
\newpage

\section{Introduction}
    As \citet{mishkin2006financial} point out, asymmetric information is one of the core issues in the existence of financial institutions. Financial intermediaries, such as commercial banks, play an important role in the financial system because they reduce transaction costs, share risk, and solve problems raised by asymmetric information. One of the most important channels of achieving this role is the effective analysis of the quality of potential corporate borrowers. Banks need to distinguish reliable borrowers from unreliable ones in order to make their decisions on corporate loans. From the banks' point of view, this reduces the losses associated with corporate defaults, while it is also beneficial for the whole economy because resources are efficiently attributed to prominent projects.\\

    \cite{altman1968financial}, \cite{beaver1966financial} and \cite{ohlson1980financial} are pioneers of using statistical models in the prediction of default. They have used financial ratios which are calculated from the balance sheet and the income statement. Their inspiring work has been widely recognized, which is proved by the fact that the method has become the standard of credit risk modeling for many financial institutions. One might doubt, however, if the phenomenon of default can be ``explained'' by the financial ratios. Intuitively, default takes place when the cash flows of a firm are no longer sustainable. The financial structure of a firm might well be the result of an upcoming default instead of being the cause of it because the firm might be obliged to sell some of its assets when it is short of cash flows. \cite{leland2006structural} distinguishes two kinds of credit risk models: structural models and statistical models (or reduced form models). According to him, the statistical model above is not directly based on firm's cash flows or values, but empirically estimates a ``jump rate'' to default. What's more, reduced form models do not allow an integrated analysis of a firm's decision to default or its optimal financial structure decisions. On the other hand, structural models, such as those proposed by \cite{black1973pricing}, \cite{merton1974pricing} and \cite{longstaff2005corporate}, associate default with the values of corporate securities, as the valuation of corporate securities depends on their future cash flows, which in turn are contingent upon the firm's operational cash flows. The diffusion models of market values of securities allow us to investigate the evolution of cash flows, and thus the default probabilities.\\

    This suggestion is insightful, but does not provide a practical approach for commercial banks vis-a-vis their corporate clients. Most small and medium-sized enterprises do not sell marketed securities. For these firms, using structural models based on corporate securities is simply impossible. Fortunately, however, commercial banks possess the information on cash flows in another way. Corporate clients not only borrow from banks but also open checking accounts in these banks. \cite{norden2010credit} demonstrate that credit line usage, limit violations, and cash inflows exhibit abnormal patterns approximately 12 months before default events. Measures of account activity substantially improve default predictions and are especially helpful for monitoring small businesses and individuals. This is another good example of economies of scale in which a bank shares information within itself to achieve better global performance.\\

    Instead of using a structural model, we choose to use some statistical learning methods which improve considerably the prediction performance compared with classical logistic regression. This choice is due to the fact that it is difficult to construct a structural model at the first stage which gives a general image and a good prediction at the same time. There is limited literature which explains the default by using checking account information. By using statistical learning methods, we can empirically tell which variables are the most important in default prediction. This can help us in the next stage construct a structural model. On the other hand, if we are only interested in prediction, a reduced form model is sufficient for our concern.\\

    However, We should underline the fact that application of machine learning methods does not eliminate the necessity of economic understanding. As we will show, the construction of meaningful economic variables is an essential preliminary step for machine learning. What's more, the ``important variables'' given by machine learning should be taken with a grain of salt. \cite{strobl1993variable} resume that variable selection in CART (classification and regression trees) is affected by characteristics other than information content, e.g. variables with more categories are preferred. To solve the problem, \cite{strobl2007unbiased} propose an unbiased split selection based on exact distribution hypothesis. As with all exact procedures, this method is computationally too intensive. \cite{hothorn2006unbiased} propose a more parsimonious algorithm, conditional classification tree (ctree), which is based on the framework of permutation test developed by \cite{strasser1999asymptotic}. What's more, unbiased random forest (conditional random forest, or cforest) is constructed based on ctree. But cforest is still too heavy to be executed for our data. Besides, it is not clear whether the unbiasedness in the sense of random forest is still valuable for other machine learning methods. That is to say, it is disputable to find an universally valuable subset of variables which contain the same level of information in any statistical method. Instead of using these computationally expensive methods, we will compare the variables selected by boosting, stepwise selection and lasso. An thorough understanding of these machine learning methods is efficient to shed light on the interpretation of model selections.\\

    We begin by introducing basic random forest and boosting, as well as some important modifications to accommodate characteristics in our data.  Section \ref{sec:account} compares three approaches of treating checking account data, illustrates the importance of economically meaningful variables and shows some particularities of machine learning methods. Section \ref{sec:merge} compares the performance of financial ratios and questionnaires with that of account data, combines the two data to achieve better prediction performance. Section \ref{sec:interpret} does three model selections, respectively based on AIC, lasso and boosting. We use the logistic regression to interpret the marginal effect of these most important variables. Section \ref{sec:conclude} concludes the article.

\section{Introduction to random forest and boosting}
\label{sec:intro_ML}
    \subsection{Classification tree}
\begin{figure}[h]
          \centering
          \includegraphics[width=\textwidth]{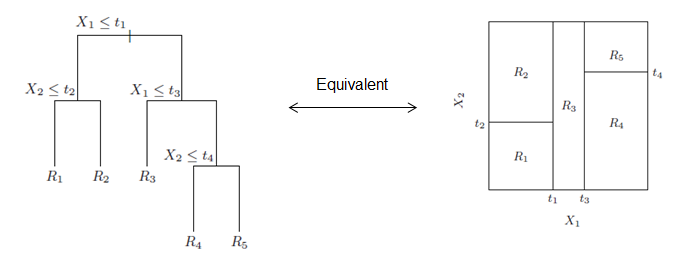}
          \caption{A simple example of classification tree}\label{fig:tree}
        \end{figure}

        For random forest and boosting, the most commonly used basic classifier is the classification tree. Suppose we want to classify a binary variable $Y$ by using two explicative variables $X_1$ and $X_2$. An example of the classification tree is given in Figure \ref{fig:tree}\footnote{Extracted from \cite{introd_SL}}. The two graphical representations are equivalent. And the tree can be represented by the form

        \begin{equation}
          \hat{f}(X)=\sum_{m=1}^{5}c_m I\{(X_1, X_2)\in R_m\} \quad \text{where $c_m \in \{0, 1\}$, $I$ is the indicator function.}
        \end{equation}

        To grow a tree, the central idea is to choose a loss function and to minimize the loss function with respect to the tree. \cite{elements_SL} and \cite{introd_SL} give a full introduction to the most important loss criteria in the context of classification trees. We use the Gini index as the loss function in our research. It should be underlined, however, that it is computationally too expensive to find a global optimal solution. Instead, in practice one uses the ``greedy algorithm'' which admits the part already constructed and searches the optimal solution based on this part. A tree grown in this way is called a CART (classification and regression tree), which was proposed by \cite{breiman1984classification} and has become the most popular tree algorithm in machine learning.\\

        The advantage of tree is obvious: it is intuitive and easy to be interpreted. Nonetheless, it generally has poor predictive power on training set and test set if the model is mildly fitted. Conversely, an overfitting with training set (or overly reduced bias) is generally not expected in machine learning. Ensemble methods, such as Random Forest and Boosting, are conceived to solve this dilemma.

    \subsection{Random Forest}

        \begin{figure}[htb]
          \centering
          \includegraphics[width=0.8\textwidth]{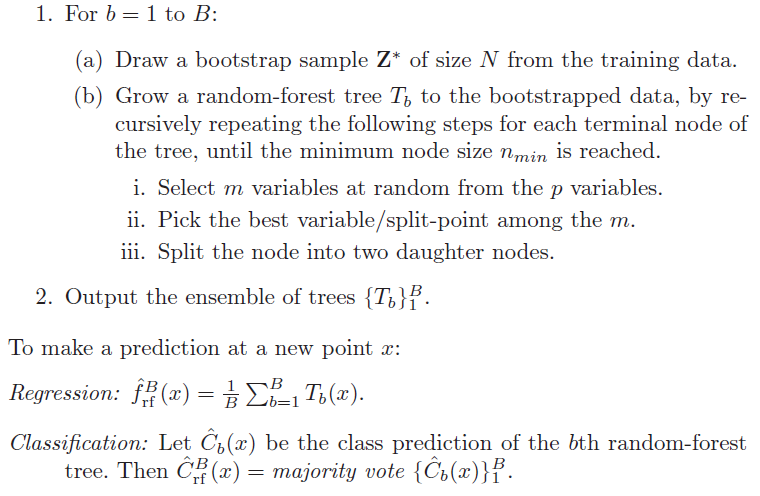}
          \caption{Algorithm of random forest}\label{fig:rf}
        \end{figure}

        Random Forest aims at reducing model variance and thus increasing prediction power on test set. Instead of growing one single tree, we plant a forest. A general description of the algorithm is given in Figure \ref{fig:rf}\footnote{Extracted from \cite{elements_SL}}. In practice, the optimal value of $m$ is around $\sqrt{p}$ for classification problem, where $p$ is the total number of variables. We can of course, use cross-validation to optimise the value of this parameter. This small value of $m$ looks strange at first sight, but it is in fact the key of random forest. In fact, for $B$ identically distributed random variables, each with variance $\sigma^2$ and positive pairwise correlation $\rho$, the variance of their average is

        \begin{equation}
          \rho \sigma^2 + \frac{1-\rho}{B}\sigma^2
        \end{equation}

        Even with large $B$ (the number of trees in the case of random forest), we still need to decrease $\rho$ to reduce the variance of average. The role of a small $m$ is to reduce the correlation $\rho$ across trees, thus decrease the model variance.\\

        However, the basic random forest works poorly for our data because it is imbalanced (fewer than $6\%$ observations defaulted). Several remedies exist for this characteristic, including weights adjustment (\cite{ting2002instance}) and stratified sampling (\cite{chen2004using}). We have adopted the stratification method which is easy to be implemented and yields satisfying results. Instead of sampling uniformly default and non-default observations for each tree in step 1.(a) (eg. sampling 2/3 observations uniformly), we take 2/3 default observations and an equal number of non-default observations. This apparently small modification leads to tremendous amelioration in confusion matrices. For a given checking account data with 30 variables, the comparison is shown in Table~\ref{tab:mat_rf}. The test AUCs are respectively $78.72\%$ and $79.87\%$.

\begin{table}[htbp]
  \centering
  \caption{Add caption}
    \begin{tabular}{|c|c|r|r|r|c|c|r|r|}
    \multicolumn{4}{c}{Training set} & \multicolumn{1}{r}{} & \multicolumn{4}{c}{Test set} \\
\cmidrule{1-4}\cmidrule{6-9}    \multicolumn{1}{|r}{} &       & \multicolumn{2}{c|}{Error rate} &       & \multicolumn{1}{r}{} &       & \multicolumn{2}{c|}{Error rate} \\
\cmidrule{3-4}\cmidrule{8-9}    \multicolumn{1}{|r}{} &       & \multicolumn{1}{l|}{Imbalanced} & \multicolumn{1}{c|}{Balanced} &       & \multicolumn{1}{r}{} &       & \multicolumn{1}{l|}{Imbalanced} & \multicolumn{1}{c|}{Balanced} \\
\cmidrule{1-4}\cmidrule{6-9}    \multirow{2}[4]{*}{True value} & \multicolumn{1}{r|}{0} & 0.057\% & 25.829\% &       & \multirow{2}[4]{*}{True value} & \multicolumn{1}{r|}{0} & 0.055\% & 25.588\% \\
\cmidrule{2-4}\cmidrule{7-9}          & \multicolumn{1}{r|}{1} & 98.861\% & 28.599\% &       &       & \multicolumn{1}{r|}{1} & 99.108\% & 27.340\% \\
\cmidrule{1-4}\cmidrule{6-9}    \multicolumn{2}{|c|}{Global} & 3.830\% & 25.930\% &       & \multicolumn{2}{c|}{Global} & 3.940\% & 25.660\% \\
\cmidrule{1-4}\cmidrule{6-9}    \end{tabular}%
  \caption{Error rates of imbalanced and balanced random forest. False negative rates are extremely high for both the training set and the test set using imbalanced random forest.In contrast, the errors rates using balanced random forest are much more reasonable.}
          \label{tab:mat_rf}
\end{table}%

    \subsection{Boosting}
        The most commonly used version of boosting is AdaBoost (\cite{freund1996experiments}). Contrary to random forest which plants decision trees in parallel, AdaBoost cultivates a series of trees. If an observation is wrongly classified in previous trees, its weight will be accentuated in latter trees until it is correctly classified. The central idea is intuitive, yet it had been purely an algorithmic notion until \cite{friedman2000additive}, who pointed out the inherent relationship between AdaBoost and additive logistic regression model:

        \begin{theorem}
          The real AdaBoost algorithm fits an additive logistic regression model by stagewise and approximate optimization of $J(F)=E[e^{-yF(x)}]$.
          \label{thr:boost_logit}
        \end{theorem}

        where additive logistic regression model is defined as having the following form for a two-class problem:

        \begin{equation}
          log\frac{P(y=1|x)}{P(y=0|x)}=\sum_{m=1}^{M}f_m(x)
        \end{equation}

        In the case of boosting trees, $f_m$ are individual trees adjusted by weights. According to Result \ref{thr:boost_logit}, boosting is by its nature an optimisation process. This insight paves the way for xgboost (Extreme Gradient Boosting by \cite{friedman2001greedy}), which searches the gradient of objective function and implements efficiently the basic idea of boosting. Moreover, the intimate relationship between boosting and logistic regression leads to some interesting results on which we will discuss later on.

    \subsection{Overfitting in machine learning}
        A model is overfitted if it suits well the training set but poorly the test set. In our research, the model performance criterion is AUC (Area Under the ROC Curve), which measures the discrimination power of a given model.
        It should be noticed that AUC is immune to imbalance in data.\\

        Some methods, like the random forest, aim at reducing the model variance, i.e., by decorrelating the training data and the model, we obtain a model which is less sensitive to data change. For example, using 30 checking account variables to explain default, we get $AUC=79.45\%$ for training set and $AUC=79.85\%$ for test set in balanced random forest. Boosting had also been considered to work in this way. But \cite{friedman2000additive} point out that boosting seems mainly a bias reducing procedure. This conclusion is coherent with our experiment. Using the same variables, we get $AUC=87.45\%$ for training set and $AUC=79.8\%$ for test set. Boosting has necessarily overfitted the model, but this feature does not undermine its ability of predicting the test set.\\

        Additional remarks should be made on parameters in machine learning methods. While it is not the major concern of this article, it is nonetheless crucial to let the machine run correctly. One important parameter is related to the complexity of model, for example, the number of candidate variables for each node splitting in random forest, the number of learning steps in boosting. Cross-validation is adopted to ensure the appropriate level of complexity and to avoid over-fitting. Appendix \ref{sec:app_param} gives an exhaustive explanation on the most important parameters in our models.

\section{Organising checking account data: Three approaches}
\label{sec:account}
    In current literature, treating checking account data does not have mature approaches as we can find for financial structure data. In the latter case, corporate finance suggests some particularly useful ratios such as working capital/total assets, retained earnings/total assets, market capitalization/total debt etc (\cite{ross2008fundamentals}). Defining new features based on checking account data becomes a central issue in our study. We have tried three approaches detailed below. They will be combined with three different statistical methods (logistic regression, random forest and boosting).
    \subsection{Variable Definition 1 (Continuous Variables Based on Economic Intuition)}
        This definition is inspired by \cite{norden2010credit}. At the end of each year, which we note time t, we define the explained variable, default, as the binary variable of going bankrupt in the next year. The explicative variables are created based on monthly account variables in the last two years. These 30 variables are listed in Appendix \ref{sec:app_var_def1} and can be classed mathematically into four categories: the difference of a characteristic (eg. balance, monthly cumulative credits) between the begin and the end of a period (one or two years); the value of this characteristic at time t; the standard deviation of this characteristic during a certain period; attributes of the firm (annual sales, sector). The basic idea is to use stock and flow variables for a complete but also concise description of a certain characteristic. Moreover, the standard deviation of, e.g. monthly cumulative credits, allows us to quantify the risk associated with unstable income.\\

        The size of firms may influence considerably the model in an undesirable way. A firm might have a higher balance than another one only because it is larger: this larger balance does not ``reflect'' a smaller probability of default. \cite{norden2010credit} have used the line of credit as the normalisation variable for the corporate clients of a German universal bank. However, this variable is not available in our research. We thus need to figure out another appropriate normalisation variable. One suggestion is to use information on the balance sheet or the income statement, such as total sales. But larger firm may open accounts in several different commercial banks, reflecting only a fragment of cash flow information in each account. There exists thus a discrepancy between the size of the account and the size of the firm. In order to capture the account size, we need a variable within the account itself which reflects the account's normal level of vitality. The average monthly cumulative credits in the last two years, responds to the defined criteria and is used to normalize the variables proportional to account size. Intuitively, monthly cumulative credits is the equivalent of total sales in the context of checking account in the sense of total resources. \\

    \subsection{Variable Definition 2 (Automatically built variables)}
        As well as in Definition 1, we still use account information in the past year to predict the default in the coming year. But the explanatory variables used in statistical methods are built in a much more ``computer science'' way. Instead of using economic intuitions above to organise raw information, we rely on automatic methods to build the model inputs. 50 variables are firstly resumed from raw monthly information, and then interact with each other using the four basic arithmetic operations. Together with some raw variables, the data set contains around 5000 variables in total. It should be noticed that these combinations are usually not intuitively interpretable. While it might be possible to give some far-fetched explanation for ``average monthly balance/cumulative number of intended violations'', it is far more difficult to interpret other variables.\\

        One might argue that the simple arithmetic interactions are not capable of exhausting possible meaningful combinations of raw information, making this approach not representative. However, it should first of all be noticed that boosting with 5000 variables is already computationally expensive for an ordinary computer. In practice, we launch the boosting for each kind of arithmetic interaction and select the most important ones according to their contributions to the Gini index. These variables are then used to run a final and lighter boosting with around 200 variables. Secondly, it is simply computationally impossible to exhaust most meaningful combinations. Suppose we want to create automatically the 30 variables in definition 1. These variables are based on more than 10 basic monthly variables (e.g. TCREDIT, monthly number of violations), i.e. more than 120 variables if we take the month into consideration. Var16 is the difference of TCREDIT between time t and t-12 (substraction of 2 variables), while var9 is the sum of monthly number of violations during one year (sum of 12 variables). This simple example shows that for a new variable, there is no limit a priori on the number of participating raw monthly variables. That is to say, any variable among the 120 variables might be included in or excluded from the combination. The number of possible forms of combination is astronomical: $2^{120}$, even if we allow only one arithmetic operation, for example the addition. Let alone other forms of operations. Thirdly, there is no reason to delimitate a priori a set of reasonable operations. The use of standard deviation for TCREDIT (var24), for example, is based on the intuition of the stability of revenue. It is not reasonable, however, to include a priori this operation, which is more complicated than sinus, cosinus or other simple functions, in the set of reasonable operations, if we investigate the question in a purely mathematical way.

    \subsection{Variable Definition 3 (Discrete Variables Based on Economic Intuition)}
        Similar to Definition 1, this definition is also economically interpretable. In contrast, we create 5 variables which are highly discretized. Four of them are binary, the fifth has three categories. These variables are listed in Appendix \ref{sec:app_var_def3}.

    \subsection{Comparison of Performance}

        \begin{table}[h]
          \centering

            \begin{tabular}{|c|l|r|r|r|r|}
            \toprule
            \multicolumn{2}{|c|}{\multirow{2}[4]{*}{}} & \multicolumn{4}{c|}{Data} \\
        \cmidrule{3-6}    \multicolumn{2}{|c|}{} & \multicolumn{2}{c|}{Definition 1} & \multicolumn{1}{c|}{Definition 2} & \multicolumn{1}{c|}{Definition 3} \\
            \midrule
            \multicolumn{2}{|c|}{Total number of variables in the definition} & \multicolumn{2}{c|}{30} & \multicolumn{1}{c|}{5000} & \multicolumn{1}{c|}{5} \\
            \midrule
            \multicolumn{2}{|c|}{} & \multicolumn{1}{l|}{Group 1} & \multicolumn{1}{l|}{Group 2} & \multicolumn{1}{l|}{Group 3} & \multicolumn{1}{l|}{Group 4} \\
            \midrule
            \multicolumn{2}{|c|}{Number of variables used in the group} & 20    & 5     & 20 (17) & 5 \\
            \midrule
            \multirow{4}[8]{*}{Test AUC} & logit & 70.87\% & 74.42\% & 74.64\% & 74.09\% \\
        \cmidrule{2-6}          & imbalanced random forest & 78.41\% & 71.96\% & 75.31\% & 46.81\% \\
        \cmidrule{2-6}          & balanced random forest & 80.02\% & 76.35\% & 76.67\% & 72.46\% \\
        \cmidrule{2-6}          & boosting & 79.66\% & 77.46\% & 78.13\% & 74.24\% \\
            \bottomrule
            \end{tabular}%
            \caption{Test AUCs of four groups of account data (3 definitions) in logit, random forest and boosting. The 20 variables in Group 1 and Group 3 are selected respectively by AIC and by variable importance in boosting. The 5 best variables in Group 2 are chosen according to variable importance in boosting. All the variables in Group 2 are included in Group 1. Among the 20 variables in Group 3, three variables are not available for most of the observations ($>50\%$) and are eliminated for random forest and for logistic regression.}
          \label{tab:perf_account}%
    \end{table}%

        The performance, measured by test AUC, is given in Table~\ref{tab:perf_account}. We have selected the 20 best variables in Group 1 and Group 3 respectively by AIC and by variable importance in boosting. The 5 best variables in Group 2 are chosen according to variable importance in boosting. Despite the difference in variable selection methods, all the variables in Group 2 are included in Group 1. Among the 20 variables in Group 3, three variables are not available for most of the observations ($>50\%$) and are eliminated for random forest and for logistic regression.\\

        We can remark that balanced random forest and boosting always outperform logistic regression, except for Group 4. (The failure of imbalanced random forest vis-a-vis balanced random forest justifies the use of stratification for imbalanced data.) This result clearly favors the application of machine learning methods in the default prediction for our data. But why does boosting exhibit the same level of performance as logit for Group 4? For this group, the logit even outforms the balanced random forest. It seems to us that discretization is the reason for this. While it is a common approach to discretize continuous variables for logistic regression because this can create a certain kind of non-linearity of a given explanatory variable within the linear framework, this will nonetheless reduce the information contained in it. The discretization is especially detrimental for imbalanced random forest. $AUC=46.81\%$ suggests a worse performance than randomly distributed classes and should be considered as a pathology. Even the balanced random forest performs worse than logistic regression. In fact, the individual trees grown in a random forest are usually very deep ($depth>1000$ with default settings for our data). The capacity of classification is thus intimately related to the number of potential splits permitted by the variables. The split of a specific node in a classification tree can be seen as an automatic discretization process. It would better to let the tree choose the splitting point by itself according to the optimisation criterion rather than fixing it a priori. As for boosting, the trees are usually shallow ($depth=5$ in our setting). As \cite{elements_SL} suggest, experiences so far indicate that $4<=depth<=8$ works well in the context of boosting, with results being fairly insensitive to particular choices in this range. In any case, it is unlikely that $depth>10$ will be required. This probably suggests that boosting relies much less heavily on the variables' ability of offering potential splits, making it less sensitive to discrete variables.\\

        In fact, using stumps ($depth=2$) is sufficiently efficient for yielding good prediction. Using all the 30 variables in Definition 1, the AUCs are respectively $79.47\%$ for $depth=2$ and $79.82\%$ for $depth=5$. (It should be remarked, however, that the optimal number of rounds validated by cross-validation is higher in the case $depth=2$. They are 2811 and 997 respectively for $depth=2$ and $depth=5$ with other parameters fixed according to Appendix \ref{sec:app_param}.) In the case of $M$ stumps, the additive logistic regression model becomes:
        \begin{equation}
          log\frac{P(y=1|x)}{P(y=0|x)}=\sum_{m=1}^{M}\alpha_m 1_{x_m<s_m} \quad \text{where } x_m\in \{x^1, x^2, ..., x^p\}\text{, $p$ is the number of explicative variables}
        \end{equation}
        Remark that $M$ is usually much larger than $p$. In the case above, the ratio M/p is about 93, which means that on average 93 dummy variables are created for each continuous explanatory variable. The linear combination of these dummies can be very good approximation of any ordinary non-linear function. With $depth>2$, the approximation is extended to multivariate functions. So the advantage of boosting over logistic regression seems to be the capacity of the former to take non-linearity into consideration. This clearly explains why boosting is mainly a bias-reducing method, as mentioned by \cite{elements_SL}.\\

         Does the out-performance of boosting and balanced random forest also imply their superiority of identifying rich data set? Comparing Group 1 and Group 3, we can remark that logit AUC is higher in Group 3, while boosting AUC is lower. If we trust in logistic regression in the case of prediction, we should conclude that Group 3 contains more information than Group 1 and that machine learning methods such as boosting are not reliable for distinguishing a rich data set from a poorer one. However, looking at Group 2, we can easily reverse this conclusion. The logit AUC in Group 2 is nearly the same as that in Group 3, while Group 2 contains apparently less information than Group 1 because all the variables in Group 2 are included in Group 1. Instead, a plausible explanation for the low logit AUC in Group 1 should be the multicollinearity between explanatory variables (\cite{introd_SL}). Boosting and random forest, in contrast, split each node by individual variable and should not be impacted by the haunting multicollinearity. With less variables (Group 2), the prediction accuracy is higher in logit. This phenomenon probably suggests that logit can not well ``digest'' rich information because of its restrictive linear form. It is thus more reliable to use AUCs of machine learning methods as a measure of information contained in the data set.\\

        The close relationship between boosting and logistic regression explains some results which may seem strange at first sight. The higher logit AUC in Group 3 compared with Group 1 should be interpreted by the model selection method: ``good variables'' in the sense of boosting should generally be ``good'' in the sense of logit. It is thus not surprising to find that 20 variables selected from about 5000 variables works better in logit than 20 variables selected from 30. On the other hand, the same variables have lower AUC in balanced random forest than in boosting ($76.67\%$ vs $78.13\%$). Does this mean that random forest is worse than boosting for predicting? If we look at Table~\ref{tab:perf_merge}, balanced random forest and boosting generally have the same prediction power. The difference between them for Group 3, as well as for Group 2, should be explained by the model selection method. These variables are selected by their contribution to boosting and naturally fit less well the random forest. These phenomena raise the question of the existence of universally valuable selection method, in the sense that the ``good'' variables are equally good for any machine learning method, not just for one or several methods which is identical or are close to the method used in the selection process. While this question is difficult to answer, we can at least conclude that variables should be defined a priori (Definition 1) based on economic intuition, instead of by a pure ``computer science'' way (Definition 2) and being selected by machine learning methods at a latter stage. Besides the bias in variable selection, automatically created variables in Group 3 also contain less information than Group 1 (AUCs of both balanced random forest and boosting are lower in Group 3). At least for our data, an economically meaningful construction of variables allows us not only to interpret marginal effects of explanatory variables but also possess information in a more concise way. The discussion can be extended to controversial epistemological discussions which are out of the scope of the article. But at least it seems to us that, as \cite{williamson2009philosophy} points out, while someone hope that machine learning can ``close the inductive loop''-i.e., automate the whole cyclic procedure of data collection, hypothesis generation, further data collection, hypothesis reformulation...- the present reality is that machine successes are achieved in combination with human expertise. \\

\section{Combination of checking account information, financial ratios and managerial information}
    \label{sec:merge}
    Traditional reduced form methods for default prediction mainly focused on financial structure of enterprises(\cite{altman1968financial}, \cite{beaver1966financial} and \cite{ohlson1980financial}), as financial structure does reflect to a large extent the solvability of enterprises, and is relatively more available than real time account information. What's more, in a reduced form method, as we merely try to match a pattern to the data (\cite{fayyad1996data}) without worrying much about causality, the problem of endogeneity is not a primary concern. But once we want to get some causal interpretation, financial structure data may suffer from endogeneity and should be carefully interpreted as a ``cause'' of credit default. On the other hand, we should remark the difference between book value and market value, and the accounting principle associated with this difference (\cite{ross2008fundamentals}). For small and medium-sized enterprises, their market values are simply not available because they usually don't sell any marketed securities, while their book values are historic and subjected to accounting manipulations.\\

    Commercial banks have both a necessity and an advantage in the analysis of credit default. The possession of corporate account information helps them to acquire a more direct and ``frank'' image of the firms' account. Not only the information may be more reliable, but also more real-time. Balance sheet and income statement are resumed once a year by firms, while checking account information can be theoretically daily. In practice, we use monthly variables as raw variables for the purpose of simplification. This allows commercial banks to supervise the solvability of corporate borrowers on a more frequent basis. Given the advantage of checking account information, we should expect a better performance of prediction based on checking account data. This is represented in the first two columns of Table~\ref{tab:perf_merge}. The AUCs based on account data in balanced random forest and boosting are significantly larger than those based on financial and management data. (One might argue that this superiority is simply due to more explicative variables. In fact, with the same number of variables (11), the boosting AUC of account data is $79.19\%$, which is nearly the same as that of 20 account variables ($79.66\%$) and significantly larger than that of financial and managerial variables ($76.17\%$)).\\

    \begin{table}[h]
      \centering

        \begin{tabular}{|c|l|r|r|r|}
        \toprule
        \multicolumn{2}{|c|}{\multirow{2}[4]{*}{}} & \multicolumn{3}{c|}{Data} \\
    \cmidrule{3-5}    \multicolumn{2}{|c|}{} & \multicolumn{1}{c|}{Definition 1} & \multicolumn{1}{c|}{Financial \& managerial} & \multicolumn{1}{c|}{Merged} \\
        \midrule
        \multicolumn{2}{|c|}{} & \multicolumn{1}{c|}{Group 1} & \multicolumn{1}{c|}{Group 5} & \multicolumn{1}{c|}{Group 6} \\
        \midrule
        \multicolumn{2}{|c|}{Number of variables used in the group} & 20    & 11    & 31 \\
        \midrule
        \multirow{4}[8]{*}{Test AUC} & logit & 70.87\% & 75.33\% & 79.55\% \\
    \cmidrule{2-5}          & imbalanced random forest & 78.41\% & 72.66\% & 83.35\% \\
    \cmidrule{2-5}          & balanced random forest & 80.02\% & 76.05\% & 83.18\% \\
    \cmidrule{2-5}          & boosting & 79.66\% & 76.17\% & 84.24\% \\
        \bottomrule
        \end{tabular}%
        \caption{The AUCs of checking account data, financial and managerial data, merged data in logit, random forest and boosting. Group 6 comes from the fusion of Group 1 and Group 5. This merged data has the best performance in balanced random forest and boosting.}
      \label{tab:perf_merge}%
    \end{table}%

    But this does not suggest the inutility of financial statements in default evaluation. In fact, if we combine Group 1 and Group 5 to yield a data with all the three components(financial ratios, questionnaires on firms and checking account data), the test AUC is the highest ever ($84.24\%$). (Once again, one might argue that the higher AUC is simply due to more variables, instead of orthogonal nature between different sources of information. This argument is refuted by our experiment: Using all the 30 variables in Definition 1, the test AUC of boosting is only $79.8\%$, which is nearly the same as that of Group 1 and significantly less than that of Group 6.) We can thus conclude that the three sources of information are complementary, which corresponds to our intuition on the real functioning of enterprises. First, The checking account information is a reflection of a firm's cash flow, which is most directly related to a firm's solvability. Second, financial ratios illustrate the firm's financial structure and its ability to earn profits. We should remark that the financial ratios we used are primarily concerned with the firm's profitability and expenses (Interest expenses, earnings before interest and taxes etc.) and are more tightly related to cash flow, which is also the case for \cite{atiya2001bankruptcy}. Third, other non-financial reasons should be taken into consideration, for example, the managerial expertise of cadres.\\

    Of course, these is not a complete list of all the factors which are related to credit default. Some macroeconomic factors, for example, can be additionally taken into account. We have observed a decreasing quarterly default rate during 2013-2014, which might be explained by decreasing interest rate in Europe during the same period. If we use data from 2009 to 2012 as training set, and that from 2013 to 2014 as test set, the statistical pattern works less well for defaults at the end of 2013 and at the beginning of 2014.

\section{Selection and interpretation of the most important variables in checking account data}
\label{sec:interpret}

\begin{figure}[h]
          \centering
          \includegraphics[width=\textwidth]{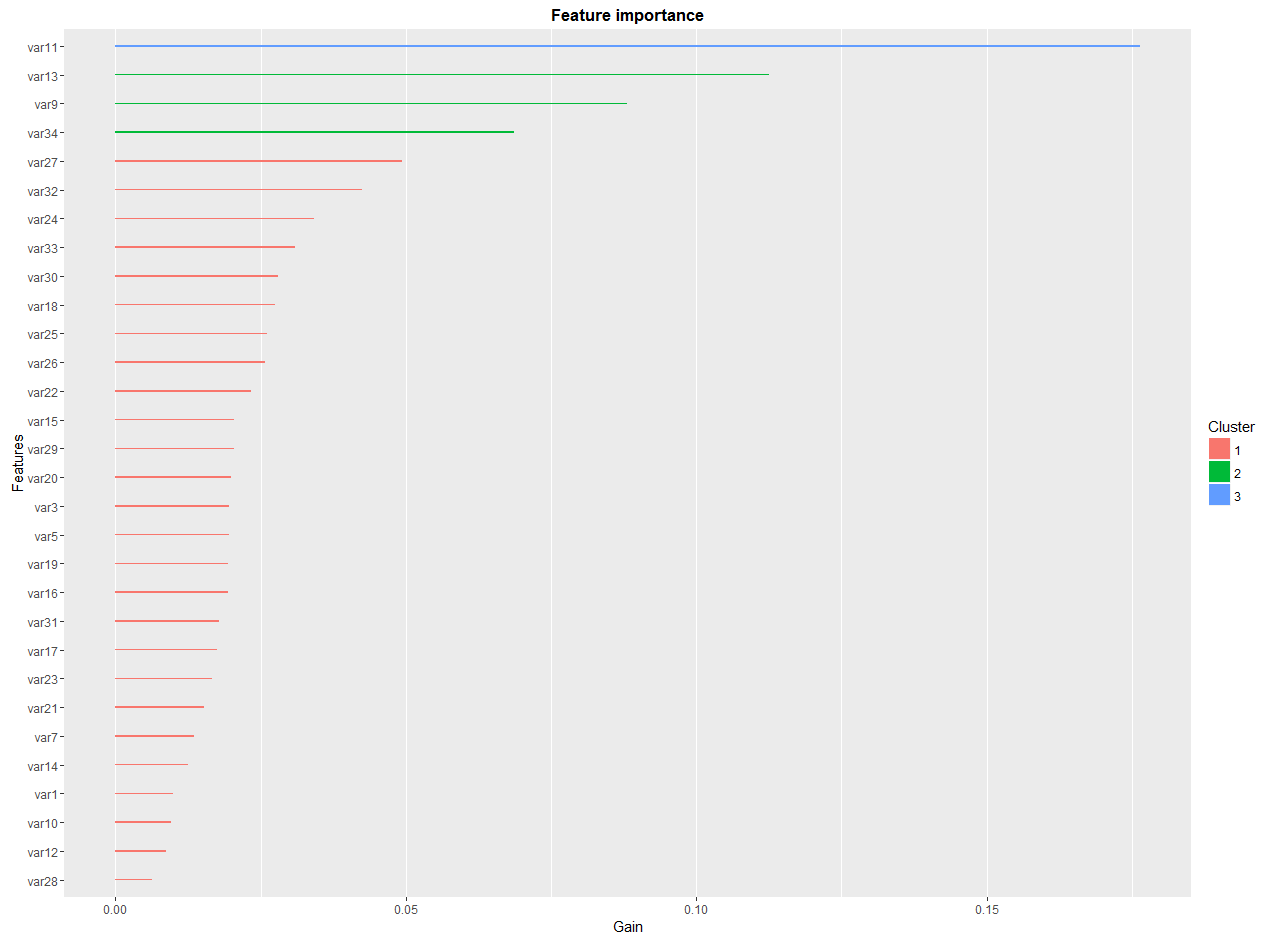}
          \caption{Variable importance of 30 variables in Definition 1 according to boosting}
          \label{fig:var_imp}
    \end{figure}

    Because of the multicollinearity problem between 30 variables in Definition 1, a variable selection process is needed in order to obtain and interpret the marginal effect of each prominent variable by logistic regression. The list of important variables in Definition 1 is shown in Figure \ref{fig:var_imp}. We can see that according to boosting, the most important variables are especially related to number of violations (var9, var11, var13) and current status (var27, var32, var33, var34). Intuitive as it be, this variable importance in the sense of boosting should be taken with a grain of salt. For example, does it mean that var10 (number of intended violations during the period $[t-23, t-12]$, ranked 29 in the importance list) is much less useful than var24 (which reflects the stability of credits during the period $[t-11,t]$, ranked 7 in the importance list)? In fact, if we draw two conditional distributions (conditioned on default) of each variable and calculate their individual AUCs which reflects their individual discriminality, the AUC of var10 ($68.68\%$) is much higher than that of var24 ($56.61\%$). This seeming paradox comes from the mechanism by which boosting attributes the variable importance. In fact, at each split in each tree, the improvement in the split criterion is the importance measure attributed to the splitting variable, and is accumulated over all trees in the boosting separately for each variable. var9 and var10 refer to the same kind of information, except that var9 concerns more recent information (period $[t-11,t]$) and naturally has a better discriminatory power than var10 ($AUC=72.68\%$ vs $AUC=68.68\%$). For each node splitting, both of the two variables are candidates. As var10 has no advantage over var9, it is rarely used for splitting and is thus considered as a ``bad'' variable. In this case, it is better to be a mediocre but irreplaceable variable than a brilliant but replaceable one. But it also shows the advantage of boosting in recognizing redundant information and eliminating them.\\

    In order to be more rigorous on variable selection, we have tried out two other different methods which are based on logistic regression: stepwise selection and lasso. For stepwise selection, AIC was used as the criterion. Forward and backward selection have generated the same 8 variables marked in Table~\ref{tab:var_select}. In order to compare between different model selection methods, we have adjusted $\lambda$ in lasso so as to yield exactly 8 non-zero coefficients. These 8 variables are also resumed in Table~\ref{tab:var_select}. Remark that there are 7 among 8 variables which are identical to those selected by AIC. Thus for our data, there is no apparent difference between stepwise and lasso in model selection. In contrast, 4 among 8 variables selected by boosting differ from that of lasso and stepwise selection! These variables all belong to the variables of current status.\\

    \begin{table}
      \centering

            \begin{tabular}{rrrlrrr}
        \toprule
        \toprule
              &       &       &       & \multicolumn{3}{c}{Selected by } \\
        \multicolumn{2}{c}{Variable category} & \multicolumn{1}{l}{Name} & Importance order in boosting & \multicolumn{1}{l}{Boosting} & \multicolumn{1}{l}{AIC} & \multicolumn{1}{l}{lasso} \\
        \midrule
        \midrule
        \multicolumn{1}{|c|}{\multirow{17}[16]{*}{Evolutions}} & \multicolumn{1}{c|}{\multirow{4}[2]{*}{Balance}} & \multicolumn{1}{l|}{\cellcolor[rgb]{ .325,  1,  .341} var1} & \multicolumn{1}{r|}{\cellcolor[rgb]{ .325,  1,  .341} 28} & \cellcolor[rgb]{ .325,  1,  .341}  & \cellcolor[rgb]{ .325,  1,  .341}  & \multicolumn{1}{r|}{\cellcolor[rgb]{ .325,  1,  .341} } \\
        \multicolumn{1}{|c|}{} & \multicolumn{1}{c|}{} & \multicolumn{1}{l|}{\cellcolor[rgb]{ .325,  1,  .341} var3} & \multicolumn{1}{r|}{\cellcolor[rgb]{ .325,  1,  .341} 19} & \cellcolor[rgb]{ .325,  1,  .341}  & \cellcolor[rgb]{ .325,  1,  .341}  & \multicolumn{1}{r|}{\cellcolor[rgb]{ .325,  1,  .341} } \\
        \multicolumn{1}{|c|}{} & \multicolumn{1}{c|}{} & \multicolumn{1}{l|}{\cellcolor[rgb]{ .325,  1,  .341} var5} & \multicolumn{1}{r|}{\cellcolor[rgb]{ .325,  1,  .341} 22} & \cellcolor[rgb]{ .325,  1,  .341}  & \cellcolor[rgb]{ .325,  1,  .341}  & \multicolumn{1}{r|}{\cellcolor[rgb]{ .325,  1,  .341} } \\
        \multicolumn{1}{|c|}{} & \multicolumn{1}{c|}{} & \multicolumn{1}{l|}{\cellcolor[rgb]{ .325,  1,  .341} var7} & \multicolumn{1}{r|}{\cellcolor[rgb]{ .325,  1,  .341} 24} & \cellcolor[rgb]{ .325,  1,  .341}  & \cellcolor[rgb]{ .325,  1,  .341}  & \multicolumn{1}{r|}{\cellcolor[rgb]{ .325,  1,  .341} } \\
    \cmidrule{2-7}    \multicolumn{1}{|c|}{} & \multicolumn{1}{c|}{\multirow{6}[6]{*}{Violations}} & \multicolumn{1}{l|}{\cellcolor[rgb]{ .573,  .804,  .863} var9} & \multicolumn{1}{r|}{\cellcolor[rgb]{ .573,  .804,  .863} 3} & \multicolumn{1}{l}{\cellcolor[rgb]{ .573,  .804,  .863} YES} & \multicolumn{1}{l}{\cellcolor[rgb]{ .573,  .804,  .863} YES} & \multicolumn{1}{l|}{\cellcolor[rgb]{ .573,  .804,  .863} YES} \\
        \multicolumn{1}{|c|}{} & \multicolumn{1}{c|}{} & \multicolumn{1}{l|}{\cellcolor[rgb]{ .922,  .945,  .871} var10} & \multicolumn{1}{r|}{\cellcolor[rgb]{ .922,  .945,  .871} 29} & \cellcolor[rgb]{ .922,  .945,  .871}  & \cellcolor[rgb]{ .922,  .945,  .871}  & \multicolumn{1}{r|}{\cellcolor[rgb]{ .922,  .945,  .871} } \\
    \cmidrule{3-7}    \multicolumn{1}{|c|}{} & \multicolumn{1}{c|}{} & \multicolumn{1}{l|}{\cellcolor[rgb]{ .573,  .804,  .863} var11} & \multicolumn{1}{r|}{\cellcolor[rgb]{ .573,  .804,  .863} 1} & \multicolumn{1}{l}{\cellcolor[rgb]{ .573,  .804,  .863} YES} & \multicolumn{1}{l}{\cellcolor[rgb]{ .573,  .804,  .863} YES} & \multicolumn{1}{r|}{\cellcolor[rgb]{ .573,  .804,  .863} } \\
        \multicolumn{1}{|c|}{} & \multicolumn{1}{c|}{} & \multicolumn{1}{l|}{\cellcolor[rgb]{ .922,  .945,  .871} var12} & \multicolumn{1}{r|}{\cellcolor[rgb]{ .922,  .945,  .871} 26} & \cellcolor[rgb]{ .922,  .945,  .871}  & \cellcolor[rgb]{ .922,  .945,  .871}  & \multicolumn{1}{r|}{\cellcolor[rgb]{ .922,  .945,  .871} } \\
    \cmidrule{3-7}    \multicolumn{1}{|c|}{} & \multicolumn{1}{c|}{} & \multicolumn{1}{l|}{\cellcolor[rgb]{ .573,  .804,  .863} var13} & \multicolumn{1}{r|}{\cellcolor[rgb]{ .573,  .804,  .863} 2} & \multicolumn{1}{l}{\cellcolor[rgb]{ .573,  .804,  .863} YES} & \multicolumn{1}{l}{\cellcolor[rgb]{ .573,  .804,  .863} YES} & \multicolumn{1}{l|}{\cellcolor[rgb]{ .573,  .804,  .863} YES} \\
        \multicolumn{1}{|c|}{} & \multicolumn{1}{c|}{} & \multicolumn{1}{l|}{\cellcolor[rgb]{ .922,  .945,  .871} var14} & \multicolumn{1}{r|}{\cellcolor[rgb]{ .922,  .945,  .871} 27} & \cellcolor[rgb]{ .922,  .945,  .871}  & \multicolumn{1}{l}{\cellcolor[rgb]{ .922,  .945,  .871} YES} & \multicolumn{1}{l|}{\cellcolor[rgb]{ .922,  .945,  .871} YES} \\
    \cmidrule{2-7}    \multicolumn{1}{|c|}{} & \multicolumn{1}{c|}{Balance Vitality} & \multicolumn{1}{l|}{\cellcolor[rgb]{ .573,  .804,  .863} var15} & \multicolumn{1}{r|}{\cellcolor[rgb]{ .573,  .804,  .863} 13} & \cellcolor[rgb]{ .573,  .804,  .863}  & \cellcolor[rgb]{ .573,  .804,  .863}  & \multicolumn{1}{r|}{\cellcolor[rgb]{ .573,  .804,  .863} } \\
    \cmidrule{2-7}    \multicolumn{1}{|c|}{} & \multicolumn{1}{c|}{\multirow{6}[6]{*}{Credits \& Debits}} & \multicolumn{1}{l|}{\cellcolor[rgb]{ .573,  .804,  .863} var16} & \multicolumn{1}{r|}{\cellcolor[rgb]{ .573,  .804,  .863} 15} & \cellcolor[rgb]{ .573,  .804,  .863}  & \cellcolor[rgb]{ .573,  .804,  .863}  & \multicolumn{1}{r|}{\cellcolor[rgb]{ .573,  .804,  .863} } \\
        \multicolumn{1}{|c|}{} & \multicolumn{1}{c|}{} & \multicolumn{1}{l|}{\cellcolor[rgb]{ .922,  .945,  .871} var17} & \multicolumn{1}{r|}{\cellcolor[rgb]{ .922,  .945,  .871} 25} & \cellcolor[rgb]{ .922,  .945,  .871}  & \cellcolor[rgb]{ .922,  .945,  .871}  & \multicolumn{1}{r|}{\cellcolor[rgb]{ .922,  .945,  .871} } \\
    \cmidrule{3-7}    \multicolumn{1}{|c|}{} & \multicolumn{1}{c|}{} & \multicolumn{1}{l|}{\cellcolor[rgb]{ .573,  .804,  .863} var18} & \multicolumn{1}{r|}{\cellcolor[rgb]{ .573,  .804,  .863} 12} & \cellcolor[rgb]{ .573,  .804,  .863}  & \cellcolor[rgb]{ .573,  .804,  .863}  & \multicolumn{1}{r|}{\cellcolor[rgb]{ .573,  .804,  .863} } \\
        \multicolumn{1}{|c|}{} & \multicolumn{1}{c|}{} & \multicolumn{1}{l|}{\cellcolor[rgb]{ .922,  .945,  .871} var19} & \multicolumn{1}{r|}{\cellcolor[rgb]{ .922,  .945,  .871} 20} & \cellcolor[rgb]{ .922,  .945,  .871}  & \cellcolor[rgb]{ .922,  .945,  .871}  & \multicolumn{1}{r|}{\cellcolor[rgb]{ .922,  .945,  .871} } \\
    \cmidrule{3-7}    \multicolumn{1}{|c|}{} & \multicolumn{1}{c|}{} & \multicolumn{1}{l|}{\cellcolor[rgb]{ .573,  .804,  .863} var20} & \multicolumn{1}{r|}{\cellcolor[rgb]{ .573,  .804,  .863} 17} & \cellcolor[rgb]{ .573,  .804,  .863}  & \cellcolor[rgb]{ .573,  .804,  .863}  & \multicolumn{1}{r|}{\cellcolor[rgb]{ .573,  .804,  .863} } \\
        \multicolumn{1}{|c|}{} & \multicolumn{1}{c|}{} & \multicolumn{1}{l|}{\cellcolor[rgb]{ .922,  .945,  .871} var21} & \multicolumn{1}{r|}{\cellcolor[rgb]{ .922,  .945,  .871} 21} & \cellcolor[rgb]{ .922,  .945,  .871}  & \cellcolor[rgb]{ .922,  .945,  .871}  & \multicolumn{1}{r|}{\cellcolor[rgb]{ .922,  .945,  .871} } \\
        \midrule
        \multicolumn{1}{|c|}{\multirow{4}[4]{*}{Risk}} & \multicolumn{1}{c|}{\multirow{2}[2]{*}{Balance stability}} & \multicolumn{1}{l|}{\cellcolor[rgb]{ .573,  .804,  .863} var22} & \multicolumn{1}{r|}{\cellcolor[rgb]{ .573,  .804,  .863} 14} & \cellcolor[rgb]{ .573,  .804,  .863}  & \cellcolor[rgb]{ .573,  .804,  .863}  & \multicolumn{1}{r|}{\cellcolor[rgb]{ .573,  .804,  .863} } \\
        \multicolumn{1}{|c|}{} & \multicolumn{1}{c|}{} & \multicolumn{1}{l|}{\cellcolor[rgb]{ .922,  .945,  .871} var23} & \multicolumn{1}{r|}{\cellcolor[rgb]{ .922,  .945,  .871} 23} & \cellcolor[rgb]{ .922,  .945,  .871}  & \cellcolor[rgb]{ .922,  .945,  .871}  & \multicolumn{1}{l|}{\cellcolor[rgb]{ .922,  .945,  .871} YES} \\
    \cmidrule{2-7}    \multicolumn{1}{|c|}{} & \multicolumn{1}{c|}{\multirow{2}[2]{*}{Credits stability}} & \multicolumn{1}{l|}{\cellcolor[rgb]{ .573,  .804,  .863} var24} & \multicolumn{1}{r|}{\cellcolor[rgb]{ .573,  .804,  .863} 7} & \multicolumn{1}{l}{\cellcolor[rgb]{ .573,  .804,  .863} YES} & \multicolumn{1}{l}{\cellcolor[rgb]{ .573,  .804,  .863} YES} & \multicolumn{1}{l|}{\cellcolor[rgb]{ .573,  .804,  .863} YES} \\
        \multicolumn{1}{|c|}{} & \multicolumn{1}{c|}{} & \multicolumn{1}{l|}{\cellcolor[rgb]{ .922,  .945,  .871} var25} & \multicolumn{1}{r|}{\cellcolor[rgb]{ .922,  .945,  .871} 11} & \cellcolor[rgb]{ .922,  .945,  .871}  & \cellcolor[rgb]{ .922,  .945,  .871}  & \multicolumn{1}{r|}{\cellcolor[rgb]{ .922,  .945,  .871} } \\
        \midrule
        \multicolumn{1}{|c|}{\multirow{7}[2]{*}{Actual}} & \multicolumn{1}{c|}{\multirow{7}[2]{*}{}} & \multicolumn{1}{l|}{var26} & \multicolumn{1}{r|}{10} &       &       & \multicolumn{1}{r|}{} \\
        \multicolumn{1}{|c|}{} & \multicolumn{1}{c|}{} & \multicolumn{1}{l|}{var27} & \multicolumn{1}{r|}{4} & \multicolumn{1}{l}{YES} &       & \multicolumn{1}{r|}{} \\
        \multicolumn{1}{|c|}{} & \multicolumn{1}{c|}{} & \multicolumn{1}{l|}{var28} & \multicolumn{1}{r|}{30} &       &       & \multicolumn{1}{r|}{} \\
        \multicolumn{1}{|c|}{} & \multicolumn{1}{c|}{} & \multicolumn{1}{l|}{var31} & \multicolumn{1}{r|}{18} &       & \multicolumn{1}{l}{YES} & \multicolumn{1}{l|}{YES} \\
        \multicolumn{1}{|c|}{} & \multicolumn{1}{c|}{} & \multicolumn{1}{l|}{var32} & \multicolumn{1}{r|}{6} & \multicolumn{1}{l}{YES} &       & \multicolumn{1}{r|}{} \\
        \multicolumn{1}{|c|}{} & \multicolumn{1}{c|}{} & \multicolumn{1}{l|}{var33} & \multicolumn{1}{r|}{8} & \multicolumn{1}{l}{YES} &       & \multicolumn{1}{r|}{} \\
        \multicolumn{1}{|c|}{} & \multicolumn{1}{c|}{} & \multicolumn{1}{l|}{var34} & \multicolumn{1}{r|}{5} & \multicolumn{1}{l}{YES} &       & \multicolumn{1}{r|}{} \\
    \cmidrule{1-2}    \multicolumn{1}{|c|}{\multirow{2}[2]{*}{Attributes}} & \multicolumn{1}{c|}{\multirow{2}[2]{*}{}} & \multicolumn{1}{l|}{var29} & \multicolumn{1}{r|}{16} &       & \multicolumn{1}{l}{YES} & \multicolumn{1}{l|}{YES} \\
        \multicolumn{1}{|c|}{} & \multicolumn{1}{c|}{} & \multicolumn{1}{l|}{var30} & \multicolumn{1}{r|}{9} &       & \multicolumn{1}{l}{YES} & \multicolumn{1}{l|}{YES} \\
        \midrule
              &       &       &       &       &       &  \\
              &       & \cellcolor[rgb]{ .325,  1,  .341}  & Information on [t-23, t] &       &       &  \\
              &       & \cellcolor[rgb]{ .573,  .804,  .863}  & Information on [t-11, t] &       &       &  \\
              &       & \cellcolor[rgb]{ .855,  .933,  .953} \textcolor[rgb]{ .573,  .816,  .314}{} & Information on [t-23, t-12] &       &       &  \\
        \end{tabular}%

        \caption{Variable selection among 30 variables of Definition 1 according to boosting, stepwise selection and lasso. Stepwise selection with AIC criterion (both forward and backward) has yielded 8 variables. For the purpose of comparison, we have chosen the 8 best variables in boosting. For lasso, we have adjusted the parameter $\lambda$ so as to yield exactly 8 non-zero coefficients.}
      \label{tab:var_select}%
    \end{table}%

    It is difficult to confirm by experiment the reason for this difference. Our intuition centers on the multicollinearity between the 4 favored variables in boosting. Table~\ref{tab:var_corr} shows the Spearman correlation between these variables\footnote{It should be more appropriate to calculate the Pearson correlation because we are interested in linear correlation in the case of logistic regression. However, this correlation is not stable with respect to manipulations such as elimination of missing values or extreme values. Spearman correlation, on the other hand, seems to be quite stable with data manipulations, which shows the advantage of tree-based methods. Tree-based methods depend on ordinal properties of variables instead of cardinal ones.}. It seems to us that because of its restrictive linear form, logit is incapable of disentangling the interweaving information contained in these variables. On the contrary, boosting seems to be able to digest this intricate information. This hypothesis is loosely confirmed by the regression results in Table~\ref{tab:logit_AIC} and \ref{tab:logit_boost}. All the coefficients of the variables selected by AIC are significantly different from zero at the $0.1\%$ level. The 4 variables favored by boosting (var27, var32, var33, var34), on the other hand, are less significant: var27 and var33 are significant at the $5\%$ level, while var32 and var34 are not significant. We should remark, however, that all the signs of these 4 variables correspond to our intuition and that var32 and var34 are not far from being significant (P values=20.89\% and 11.23\% respectively). This is a common syndrome of multicollinearity because it increases the variances of related estimated coefficients and renders the coefficients insignificantly different from zero.\\

    \begin{table}
      \centering

        \begin{tabular}{lrrrr}
          & \multicolumn{1}{l}{var27} & \multicolumn{1}{l}{var32} & \multicolumn{1}{l}{var33} & \multicolumn{1}{l}{var34} \\
    var27 & \cellcolor[rgb]{ 1,  .753,  0} 100.00\% & \cellcolor[rgb]{ 1,  .847,  .392} -18.39\% & \cellcolor[rgb]{ 1,  .808,  .224} 77.80\% & \cellcolor[rgb]{ 1,  .773,  .075} 92.58\% \\
    var32 & \cellcolor[rgb]{ 1,  .847,  .392} -18.39\% & \cellcolor[rgb]{ 1,  .753,  0} 100.00\% & \cellcolor[rgb]{ 1,  .976,  .902} 9.97\% & \cellcolor[rgb]{ 1,  .753,  0} -30.37\% \\
    var33 & \cellcolor[rgb]{ 1,  .808,  .224} 77.80\% & \cellcolor[rgb]{ 1,  .976,  .902} 9.97\% & \cellcolor[rgb]{ 1,  .753,  0} 100.00\% & \cellcolor[rgb]{ 1,  .835,  .325} 67.68\% \\
    var34 & \cellcolor[rgb]{ 1,  .773,  .075} 92.58\% & \cellcolor[rgb]{ 1,  .753,  0} -30.37\% & \cellcolor[rgb]{ 1,  .835,  .325} 67.68\% & \cellcolor[rgb]{ 1,  .753,  0} 100.00\% \\
    \end{tabular}%

         \caption{Spearman correlation between var27, var32, var33, var34}
      \label{tab:var_corr}%
    \end{table}%

    \begin{table}
      \centering

        \begin{tabular}{lccccrrr}
    \cmidrule{1-5}          & \multicolumn{1}{l}{Coefficient} & \multicolumn{1}{l}{Standard deviation} & \multicolumn{1}{l}{P value} & \multicolumn{1}{l}{Significance} &       &       &  \\
    \cmidrule{1-5}    (Intercept) & -6.670e-01 & 1.294e-01 & 2.55e-07 & ***   &       &       &  \\
        var9  & 2.060e-03 & 1.469e-04 & $<$2e-16 & ***   &       & \multicolumn{2}{c}{Significance levels} \\
        var11 & -1.626e-03 & 1.915e-04 & $<$2e-16 & ***   &       & \multicolumn{1}{l}{***} & 0.1\% \\
        var13 & -2.129e+00 & 1.003e-01 & $<$2e-16 & ***   &       & \multicolumn{1}{l}{**} & 1\% \\
        var14 & -1.048e+00 & 1.080e-01 & $<$2e-16 & ***   &       & \multicolumn{1}{l}{*} & 5\% \\
        var24 & 1.745e-01 & 2.739e-02 & 1.88e-10 & ***   &       & \multicolumn{1}{l}{.} & 10\% \\
        var29 & 1.508e-01 & 1.842e-02 & 2.64e-16 & ***   &       &       &  \\
        var30 & -1.649e-08 & 2.533e-09 & 7.48e-11 & ***   &       &       &  \\
        var31 & -1.111e-01 & 2.005e-02 & 2.98e-08 & ***   &       &       &  \\
        \end{tabular}%
        \caption{Logistic regression using variables selected by stepwise selection}
      \label{tab:logit_AIC}%
    \end{table}%

    \begin{table}
      \centering

        \begin{tabular}{lccccrrr}
    \cmidrule{1-5}          & \multicolumn{1}{l}{Coefficient} & \multicolumn{1}{l}{Standard deviation} & \multicolumn{1}{l}{P value} & \multicolumn{1}{l}{Significance} &       &       &  \\
    \cmidrule{1-5}    (Intercept) & -1.128e+00 & 7.204e-02 & <2e-16 & ***   &       &       &  \\
        var9  & 1.433e-03 & 1.146e-04 & $<$2e-16 & ***   &       & \multicolumn{2}{c}{Significance levels} \\
        var11 & -7.558e-04 & 1.587e-04 & 1.92e-06 & ***   &       & \multicolumn{1}{l}{***} & 0.1\% \\
        var13 & -2.424e+00 & 7.207e-02 & $<$2e-16 & ***   &       & \multicolumn{1}{l}{**} & 1\% \\
        var24 & 1.955e-01 & 2.008e-02 & $<$2e-16 & ***   &       & \multicolumn{1}{l}{*} & 5\% \\
        var27 & -2.493e-03 & 1.186e-03 & 0.0355 & *     &       & \multicolumn{1}{l}{.} & 10\% \\
        var32 & 5.308e-05 & 4.224e-05 & 0.2089 &       &       &       &  \\
        var33 & -1.534e-04 & 6.869e-05 & 0.0255 & *     &       &       &  \\
        var34 & 1.149e-04 & 7.246e-05 & 0.1128 &       &       &       &  \\
        \end{tabular}%
        \caption{Logistic regression using variables selected by boosting}
      \label{tab:logit_boost}%
    \end{table}%

    Returning on the regression in Table~\ref{tab:logit_AIC}, several insights can be gained from the marginal effect of these variables. First, var9, var11 and var13 are always the best variables in any method that we have used (also valid for balanced random forest, of which we haven't presented the variable selection). These variables concern intended or rejected violations of credit line. The negative sign for var11 (number of rejected violations) should not be regarded as counter-intuitive, because of the presence of var9 (number of intended violations) and its positive coefficient which is larger than that of var11 in absolute value. This suggests that larger number of violations, whether rejected or not, indicates a higher probability of default. We have used the amount of violations instead of the number of violations to construct var13, in order to capture more precisely the confidence on each client given by the bank advisor. This variable seems to work particularly well, in the sense that the same variable on the previous year, var14, is also included by stepwise selection and by lasso. This suggests that front line staff have acquired some important experiences and intuitions in distinguishing solvable clients from insolvable ones. These experiences may be hard to be formally formulated, but are truly valuable and should be paid attention to. Second, the risk of default is intimately related to the risk of income. As var24 (standard deviation of cumulative monthly credits) shows, the more the income is unstable, the more the firm is likely to default. Var31 (cumulative monthly credits at month t) is also related to credit and decreases the default probability by having more income. Credits, rather than debits, may be considered more seriously as the source of default. \cite{norden2010credit} point out that there exists a very strong correlation between debits and credits and that the latter should be considered as the constraint of the former. Increase in expenses might be direct reason for default, but income decrease or instability may be more fundamental. Third, different economic sectors clearly have different default rate. We have constructed var29 (sector) by using a theorem in \cite{shih2001selecting}. See Appendix~\ref{sec:thm_shih} for the details of the theorem.This theorem allows us to transform a categorical variable into a discrete numeric variable for classification trees. The corresponding numeric values of sectors are shown in Table~\ref{tab:sectors}. Higher values are associated with higher average default rate. This is also validated by the logistic regression in Table~\ref{tab:logit_AIC}. Fourth, larger firms are less likely to default. They are more mature than startups. Commercial banks have reason to be unwilling to lend money to startups, who in some cases might need to search investment from venture capitals or angel investors.

    \begin{table}
      \centering

        \begin{tabular}{lrrrrr}
        Sector & \multicolumn{1}{l}{Agriculture} & \multicolumn{1}{l}{Service} & \multicolumn{1}{l}{Commerce} & \multicolumn{1}{l}{Industry} & \multicolumn{1}{l}{Construction} \\
        \midrule
        Numeric value & 1     & 2     & 3     & 4     & 5 \\
        \end{tabular}%
        \caption{Transformation of sector variable to numeric variable according to average default}
      \label{tab:sectors}%
    \end{table}%

\section{conclusion}
\label{sec:conclude}
    We have investigated the relationship between corporate checking account and credit default and shown that account information outperforms traditionally used financial ratios in predicting the default for our data sets. This result aligns with our understanding of default as a phenomenon of liquidity. Checking account information reflects a more direct and real-time status of the firm's cash flow and is a privilege of commercial banks when the firm's market value is not available. Banks can exploit economies of scale and use information on the firms' checking account to make reasonable decisions on corporate loans. Despite the importance of this subject, there is currently little literature except \cite{norden2010credit}, \cite{mester2007transactions} and \cite{jimenez2009empirical}. Inspired by their work, we have investigated a broader range of explicative variables and systematically compared the performance of different data sets by statistical learning methods. We have shown that these methods, together with the AUC criterion, are more accurate and reliable approaches to measure the information contained in data sets than logistic regression.  While the latter often suffers from multicollinearity, machine learning methods such as random forest and boosting separately make use of these variables and are capable of disentangling intricate information. By using random forest and boosting, we have significantly increased the prediction accuracy. Tree-based methods have other advantages such as being immune to extreme values.\\

    We should remark particularly, however, that successful statistical learning process is achieved with human expertise. Meaningful economic variables must be first of all created based on raw checking account information, just as pioneers on corporate finance have created financial ratios based on balance sheet and income statement. We also need to normalise these variable so as to eliminate the effect of account size. As we have shown, it is technically  not possible (and epistemologically unacceptable for some) to create explicative variables which contain the same level of concise information simply by automated program. The 30 variables created by Definition 1 need to be perfected by eliminating about one half less useful variables and adding other potential important indicators. But even at this early stage, the importance of human expertise in financial study is illustrated.\\

    Financial ratios and managerial questionnaire are nonetheless still important in predicting credit default. By combining them with checking account data, the model has the best prediction performance and outperforms any other model with only one single data. This suggests a certain kind of orthogonality between the information of different data sets: the financial structure, profitability, and managerial experience should be considered in parallel with checking account information in a reduced form model.\\

    By careful approaches of model selection, we have shown some particularities of boosting in selecting important variables. We have used the 8 most important variables given by stepwise selection to gain intuitions on the mechanism of default. Violations of credit line, whether rejected or not, are particularly good indicators of upcoming default. Moreover, front line advisors seem to have notable experience in distinguishing acceptable violations, which is reflected in the percentage of permitted amount of violations. While the default is at first sight due to excessive expenses, \cite{norden2010credit} and us have focused on the importance of credits. Low level of income, as well as instability of income, increases significantly the default rate.\\

    Our research have adopted rigorous statistical methods to obtain a well-performed prediction model based on checking account and to identify key indicators in this data by an inductive methodology. We had a thorough discussion on the mechanisms of these methods which have significant implications on the results. This has enriched the scarce literature on this topic and can provide suggestions to banks on their decision of corporate loans. Further research may try to identify other key factors in checking account information or construct a structural model for credit default of small and medium sized enterprises.


\newpage
\appendix

\section{Variable Definition 1}
\label{sec:app_var_def1}

    For simplicity, variables are abbreviated according to Table~\ref{tab:var_abbr}.
    \begin{table}[!h]
      \centering

        \begin{tabular}{ll}
        Abbreviation & Explanation \\
        \midrule
        \midrule
        MIN\_BAL & monthly min account balance\\
        MAX\_BAL & monthly max account balance\\
        MEAN\_BAL & monthly average account balance\\
        MEAN\_CRBAL & monthly average credit balance\\
        MEAN\_DBBAL & monthly average debit balance\\
        TCREDIT & monthly total credits \\
        TDEBIT & monthly total debits \\
        INT\_CNVIOL & cumulative number of intended violations from the beginning of the year \\
        REJ\_CNVIOL & cumulative number of rejected violations from the beginning of the year \\
        INT\_CAVIOL & cumulative amount of intended violations from the beginning of the year \\
        REJ\_CAVIOL & cumulative amount of rejected violations from the beginning of the year \\
        $MEAN\_TCREDIT_t$ & mean TCREDIT during the period [t-23, t], used for nomalisation\\
        \end{tabular}%
        \caption{Variable Abbreviations}
      \label{tab:var_abbr}%
    \end{table}%

    The 30 variables defined in Definition 1 are built by applying the operations in Table~\ref{tab:operations}. Their definition formulas are shown in Table~\ref{tab:def1}.\footnote{One might wonder why the variables are not nominated from 1 to 30. This is purely a historical problem: we have done a first version of 30 variables before modifying them to get the second version that we see right now.}

    \begin{table}[!h]
      \centering

        \begin{tabular}{rl}
    \multicolumn{1}{l}{Operation} & \multicolumn{1}{l}{Meaning} \\
    \midrule
    \midrule
         $X_t$ & Value of X at month t\\
         $\Delta X_t$ & $X_t-X_{t-11}$ \\
         $\Delta\Delta X_t$ & $X_t-X_{t-23}$  \\
         $mean_t(X)$ & Mean of X during the period [t-11, t] \\
         $sd_t(X)$ & Standard Deviation of X during the period [t-11, t] \\
    \end{tabular}%
        \caption{Operations for creating variables}
      \label{tab:operations}%
    \end{table}%

    \begin{sidewaystable}[htbp]
      \centering

        \begin{tabular}{|c|c|l|l|l|l|}
        \midrule
        \midrule
        \multicolumn{2}{c}{Variable category} & \multicolumn{1}{l}{Name} & \multicolumn{1}{l}{Numerator} & \multicolumn{1}{l}{Denominator} & \multicolumn{1}{l}{Use normalization} \\
        \midrule
        \midrule
        \multirow{17}[16]{*}{Evolutions} & \multirow{4}[2]{*}{Balance} & var1 & $\Delta\Delta MIN\_BAL_t$   &   $MEAN\_TCREDIT_t$    &  YES \\
              &       & var3 & $\Delta\Delta MEAN\_BAL_t$ & $MEAN\_TCREDIT_t$ & YES\\
              &       & var5 & $\Delta\Delta MEAN\_CRBAL_t$ & $MEAN\_TCREDIT_t$ & YES\\
              &       & var7 & $\Delta\Delta MEAN\_DBBAL_t$ & $MEAN\_TCREDIT_t$ & YES\\
    \cmidrule{2-6}          & \multirow{6}[6]{*}{Violations} & var9 & $\Delta INT\_CNVIOL_t$ & & NO\\
              &       & var10 & $\Delta INT\_CNVIOL_{t-12}$ & & NO\\
    \cmidrule{3-6}          &       & var11 & $\Delta REJ\_CNVIOL_t$ & & NO\\
              &       & var12 & $\Delta REJ\_CNVIOL_{t-12}$ & & NO\\
    \cmidrule{3-6}          &       & var13 & $\Delta INT\_CAVIOL_t-\Delta REJ\_CAVIOL_t$ & $\Delta INT\_CAVIOL_t$ & NO\\
              &       & var14 & $\Delta INT\_CAVIOL_{t-12}-\Delta REJ\_CAVIOL_{t-12}$ & $\Delta INT\_CAVIOL_{t-12}$ & NO\\
    \cmidrule{2-6}          & Balance Vitality & var15 & $\Delta (MAX\_BAL-MIN\_BAL)_t$ & $MEAN\_TCREDIT_t$ & YES\\
    \cmidrule{2-6}          & \multirow{6}[6]{*}{Credits \& Debits} & var16 & $\Delta TCREDIT_t$ & $MEAN\_TCREDIT_t$ & YES\\
              &       & var17 & $\Delta TCREDIT_{t-12}$ & $MEAN\_TCREDIT_t$ & YES\\
    \cmidrule{3-6}          &       & var18 & $\Delta TDEBIT_t$ & $MEAN\_TCREDIT_t$ & YES\\
              &       & var19 & $\Delta CDECIT_{t-12}$ & $MEAN\_TCREDIT_t$ & YES\\
    \cmidrule{3-6}          &       & var20 & $\Delta (TCREDIT/TDEBIT)_t$ & & NO\\
              &       & var21 & $\Delta (TCREDIT/TDEBIT)_{t-12}$ & & NO\\
        \midrule
        \multirow{4}[4]{*}{Risk} & \multirow{2}[2]{*}{Balance stability} & var22 & $sd_t(MEAN\_BAL)$ & $mean_t(MEAN\_BAL)$ & NO\\
              &       & var23 & $sd_{t-12}(MEAN\_BAL)$ & $mean_{t-12}(MEAN\_BAL)$ & NO\\
    \cmidrule{2-6}          & \multirow{2}[2]{*}{Credits stability} & var24 & $sd_t(TCREDIT)$ & $mean_t(TCREDIT)$ & NO\\
              &       & var25 & $sd_{t-12}(TCREDIT)$ & $mean_{t-12}(TCREDIT)$ & NO\\
        \midrule
        \multirow{7}[2]{*}{Actual} & \multirow{7}[2]{*}{} & var26 & $MEAN\_BAL_t$ &  $MEAN\_TCREDIT_t$ & YES\\
              &       & var27 & $MEAN\_CRBAL_t$ &  $MEAN\_TCREDIT_t$ & YES\\
              &       & var28 & $MEAN\_DBBAL_t$ &  $MEAN\_TCREDIT_t$ & YES\\
              &       & var31 & $TCREDIT_t$ &  $MEAN\_TCREDIT_t$ & YES\\
              &       & var32 & $TDEBIT_t$ &  $MEAN\_TCREDIT_t$ & YES\\
              &       & var33 & $MIN\_BAL_t$ &  $MEAN\_TCREDIT_t$ & YES\\
              &       & var34 & $MAX\_BAL_t$ &  $MEAN\_TCREDIT_t$ & YES\\
        \midrule
        \multirow{2}[2]{*}{Attributes} & \multirow{2}[2]{*}{} & var29 & sector & & NO\\
              &       & var30 & total sales & & NO\\
        \bottomrule
        \end{tabular}%
        \caption{30 variables defined in Definition 1. ``Use normalisation'' refers to the normalisation by $MEAN\_TCREDIT_t$.}
      \label{tab:def1}%
    \end{sidewaystable}%

\newpage
\section{Variable Definition 3}
\label{sec:app_var_def3}

    The 5 discrete variables are defined in Table~\ref{tab:def3}.

    \begin{table}[htbp]
      \centering

        \begin{tabular}{c|l}
        \toprule
        \toprule
        \multicolumn{1}{c}{Variable before discretization} & \multicolumn{1}{l}{Discrete classes} \\
        \midrule
        \midrule
        \multirow{2}[4]{*}{sum of MEAN\_CRBAL during [t-2, t]} &  $<=a$\texteuro\\
    \cmidrule{2-2}          &  $>a$\texteuro\\
        \midrule
        \multirow{3}[6]{*}{\begin{tabular}{@{}c@{}}sum of monthly intended number of violations S1,\\ and of monthly rejected number of violations S2,\\ during [t-2, t]\end{tabular} } &  $S1=0$\\
    \cmidrule{2-2}          &  $S1>0 \& S2=0 $ \\
    \cmidrule{2-2}          &  $S1>0 \& S2>0 $ \\
        \midrule
        \multirow{2}[4]{*}{existence of unpayed loan during [t-11, t]} &  NO\\
    \cmidrule{2-2}          &  YES\\
        \midrule
        \multirow{2}[4]{*}{$MEAN\_CRBAL_t/MEAN\_CRBAL_{t-1}$} &  $<b$\\
    \cmidrule{2-2}          &  $>=b$\\
        \midrule
        \multirow{2}[4]{*}{\begin{tabular}{@{}c@{}}history of relationship \\  with the bank (years)\end{tabular}} &  $<c$ years\\
    \cmidrule{2-2}          &  $>=c$ years\\
        \bottomrule
        \end{tabular}%
        \caption{5 variables defined in Definition 3. The exact values of $a$, $b$ and $c$ are not presented because of confidential agreement.}
      \label{tab:def3}%
    \end{table}%

\section{Parameters in Random Forest and Boosting}
\label{sec:app_param}
    In our research, the parameters of random forest and boosting are set in the following way:

    \begin{itemize}
      \item Random Forest (R package RandomForest)
        \begin{itemize}
          \item The number of candidate variables for node splitting (mtry): For a classification problem, $\sqrt{p}$ is the ``standard'' choice. We can also use cross-validation for determining the value.
          \item Balanced random forest: use the parameter ``sampsize'' for stratified sampling. If the forest is well balanced, the minimal number of observations in each node (``nodesize'') should not greatly influence the prediction power.
        \end{itemize}

      \item Boosting (R package xgboost)
        \begin{itemize}
          \item Choose the number of rounds (``nrounds'') by the cross-validation.
          \item Choose a sufficiently small number for the shrinkage parameter (``eta''). For our data, the performance is stable when $eta<0.1$. $eta=0.01$ is used in our program.
          \item Maximum depth of each tree (``max\_depth''): between 4 and 8. We have used $max_depth=5$.
          \item The proportion of observations used for each tree (``subsample'') does not influence greatly the prediction performance. We have used $subsample=0.5$.
        \end{itemize}
    \end{itemize}

\section{Theorem for transforming sector variable to discrete numeric variable}
\label{sec:thm_shih}
    \begin{theorem}
      Suppose there are two classes, class 1 and class 2. Let $X$ be a categorical variable taking values on $\{1, 2,..., L\}$ where the categories are in increasing $p(1|X=i)$ values. If $\phi$ is a concave function, then one of the $L-1$ splits, $X\in \{1, 2,..., l\}$ where $1<=l<L$, minimizes $p_{Left}\phi (p_{1Left})+p_{Right}\phi (p_{1Right})$.
    \end{theorem}

\end{document}